\documentclass[lettersize,journal]{IEEEtran}
\usepackage{amsmath,amsfonts}
\usepackage{algorithmic}
\usepackage{algorithm}
\usepackage{array}
\usepackage[caption=false,font=normalsize,labelfont=sf,textfont=sf]{subfig}
\usepackage{textcomp}
\usepackage{stfloats}
\usepackage{url}
\usepackage{verbatim}
\usepackage{graphicx}
\usepackage{cite}
\usepackage{bm}
\usepackage{amssymb}
\usepackage{multicol} 
\usepackage{caption}
\usepackage{booktabs}
\usepackage{multirow}

\begin{document}

\title{PD-Based and SINDy Nonlinear Dynamics Identification of UAVs for MPC Design}

\author{Bryan S. Guevara, Viviana Moya, Daniel C. Gandolfo, Juan M. Toibero
\thanks{Manuscript received 1 October 2023; accepted 13 January 2024. Date of publication 31 January 2024; date of current version 9 February 2024. This letter was recommended for publication by Associate Editor M. W. Mueller and Editor G. Loianno upon evaluation of the reviewers’ comments. This work was supported in part by the PhD program in Control System Engineering, UNSJ - INAUT.}
\thanks{Bryan S. Guevara, Daniel C. Gandolfo and Juan M. Toibero are with the Instituto de Automática (INAUT), Universidad Nacional de San Juan-CONICET, San Juan J5400, Argentina (e-mail: bguevara@inaut.unsj.edu.ar; dgandolfo@inaut.unsj.edu.ar, mtoibero@inaut.unsj.edu.ar).}
\thanks{Viviana Moya is with the Facultad de Ingeniería, Universidad Internacional del Ecuador, Quito, 180101, Ecuador (e-mail: vivianamoya@uide.edu.ec).}
\thanks{Digital Object Identifier 10.1109/XXX.XXXX.XXXXXXX}}

\markboth{IEEE Latin America Transactions,~Vol.~XX, No.~X, August~XXXX}%
{Shell \MakeLowercase{\textit{et al.}}: A Sample Article Using IEEEtran.cls for IEEE Journals}

\IEEEpubid{0000--0000/00\$00.00~\copyright~2021 IEEE}

\maketitle

\begin{abstract}
This paper presents a comprehensive approach to nonlinear dynamics identification for UAVs using a combination of data-driven techniques and theoretical modeling. Two key methodologies are explored: Proportional-Derivative (PD) approximation and Sparse Identification of Nonlinear Dynamics (SINDy). The UAV dynamics are first modeled using the Euler-Lagrange formulation, providing a set of generalized coordinates. However, platform constraints limit the control inputs to attitude angles, and linear and angular velocities along the z-axis. To accommodate these limitations, thrust and torque inputs are approximated using a PD controller, serving as the foundation for nonlinear system identification. In parallel, SINDy, a data-driven method, is employed to derive a compact and interpretable model of the UAV dynamics from experimental data. Both identified models are then integrated into a Model Predictive Control (MPC) framework for accurate trajectory tracking, where model accuracy, informed by data-driven insights, plays a critical role in optimizing control performance. This fusion of data-driven approaches and theoretical modeling enhances the system's robustness and adaptability in real-world conditions, offering a detailed analysis of the UAV's dynamic behavior.
\end{abstract}


\IEEEPARstart{U}{nmanned} Aerial Vehicles (UAVs) have gained significant attention in various fields, including environmental monitoring, disaster response, industrial inspections, and package delivery. As UAV platforms continue to evolve, the demand for robust control systems that ensure stability and precision in dynamic environments becomes increasingly critical. Accurate modeling of UAV dynamics is essential for the development of advanced control strategies, such as Model Predictive Control (MPC) and other optimization-based approaches \cite{Nguyen2021, Shi2021, Schwenzer2021}.

Traditional control methods often rely on simplified linear models of UAV dynamics. While these models are useful for basic flight control, they fail to capture the nonlinear behavior inherent in UAVs, which becomes particularly evident during aggressive maneuvers or when operating in dynamic environments. UAVs exhibit significant nonlinearities due to aerodynamic effects, thrust vectoring, and coupling between translational and rotational dynamics \cite{Nwaforo2024}. Therefore, the development of models that accurately reflect these dynamics is essential for improving control precision and stability.

In response to these challenges, control engineers have progressively shifted towards more sophisticated modeling techniques that better account for the nonlinearities present in UAV systems. Early methods, such as Proportional-Derivative (PD) control, offered a practical and computationally simple solution for basic flight stabilization, but they were limited in their ability to handle more complex flight conditions. As UAV missions have expanded in scope and complexity, the need for more advanced control algorithms, capable of dealing with nonlinear dynamics, multivariable interactions, and external disturbances, has become evident. This has led to the exploration of optimization-based strategies, such as MPC, and the increasing integration of data-driven methods to improve model accuracy.

In recent years, data-driven approaches have emerged as effective methods for identifying nonlinear dynamics system, leveraging real-world data to develop models that better capture the system's complex behaviors \cite{Chee2022, Torrente2021DataDrivenMF, Brunton2019 }. These methods offer a way to bypass some of the limitations associated with purely theoretical models, particularly when the physical properties of the UAV or its operational environment are difficult to quantify. Among these approaches, Sparse Identification of Nonlinear Dynamics (SINDy) has proven particularly useful in deriving compact, interpretable models of nonlinear systems \cite{sindy_fasel_2021, sparse_kaiser_2017}. SINDy operates under the assumption that many physical systems can be described by a small number of active terms, allowing for sparsity in the identification of governing equations, even in the presence of nonlinearities \cite{datadriven_manaa_2024}.

\IEEEpubidadjcol
The key advantage of SINDy lies in its ability to distill complex, high-dimensional data into simplified models that retain the essential dynamics of the system \cite{Kaiser2018}, or conversely, to effectively handle low-data scenarios while still capturing the system's core behaviors \cite{Quade2018}. By applying sparse regression techniques, SINDy identifies only the most relevant terms needed to describe the system, reducing the complexity of the model while preserving accuracy\cite{Brunton2016}. This makes it particularly suitable for real-time control applications, where computational efficiency is a priority. SINDy has been successfully applied in fields such as fluid dynamics and robotics, providing an efficient method for discovering governing equations from experimental data \cite{datadriven_fonzi_2020}. For UAVs, SINDy offers a way to derive accurate dynamic models from limited data while maintaining computational efficiency, which is crucial in applications where obtaining large datasets may be impractical or expensive \cite{sparse_quade_2018}.

Moreover, the use of data-driven models like SINDy addresses one of the key challenges in UAV control: the ability to adapt to changing conditions in real time. UAVs often operate in unpredictable environments, where factors such as wind, turbulence, and varying payloads introduce disturbances that are difficult to model a priori. Traditional model-based control methods struggle to cope with these uncertainties, as they rely on predefined models that may not account for all external variables. In contrast, SINDy allows for the incorporation of real-time data to continuously refine the dynamic model, enhancing the system's adaptability and robustness \cite{datadriven_kaiser_2019}. 

Despite these advances, accurately modeling UAV dynamics remains challenging due to the inherent nonlinearities and external disturbances, such as wind, payload variations, and sensor noise. These factors complicate the development of a single model that can account for all operational conditions \cite{datadriven_kaiser_2019}. Furthermore, UAV dynamics exhibit strong coupling between translational and rotational movements, making it difficult to separate individual effects and model them independently. Data-driven methods, such as SINDy, offer a balance between capturing complex system behaviors and maintaining computational efficiency, which is essential for real-time control applications \cite{datadriven_fonzi_2020, sindyrl_zolman_2024}.

Once the UAV dynamics are identified using a data-driven approach like SINDy, these models can be integrated into an MPC framework for tasks such as trajectory tracking and path following. MPC is well-suited for UAV applications because it handles multivariable systems, enforces input and state constraints, and optimizes control performance over a prediction horizon. The success of MPC, however, relies heavily on the precision of the dynamic model. Poorly identified models can lead to degraded tracking performance or system instability \cite{sindy_fasel_2021}. The integration of accurate models within the MPC framework ensures that UAVs can operate efficiently in both structured and unstructured environments, enabling them to perform complex maneuvers while maintaining safety and reliability.

Both the PD-approximated and SINDy-identified models offer distinct advantages within the MPC framework. The PD model provides a simple, computationally efficient approximation for control inputs, making it suitable for real-time applications. Meanwhile, the SINDy model captures more of the system’s nonlinearities, improving trajectory tracking accuracy, especially in the presence of external disturbances \cite{datadriven_manaa_2024}. The ability to combine these two approaches provides a flexible and powerful toolset for UAV control, enabling precise and robust performance across a wide range of operational scenarios.

\subsection{Contributions and Paper Outline}

This paper presents a comprehensive approach to UAV dynamics identification using PD approximation and SINDy. UAV dynamics are first modeled using the Euler-Lagrange formulation to derive the system’s equations of motion. The PD approximation is applied to model the system inputs, creating a foundational model for control. In parallel, SINDy is used to derive a sparse, nonlinear model from experimental data. Both models are integrated into an MPC framework, and their performance is evaluated through simulations and experimental validation.

The remainder of the paper is structured as follows: Section II covers the UAV dynamic modeling process, including the Euler-Lagrange formulation and PD approximation. Section III introduces the SINDy methodology for UAV dynamics identification. Section IV discusses the integration of both models into the MPC framework and presents simulation results. Finally, Section V concludes the paper and outlines directions for future work.

\section{Dynamic Model Based on Euler-Lagrange}

\begin{figure}[H]
\centering
\includegraphics[clip, trim=5cm 4cm 6cm 2cm, width=1\linewidth]{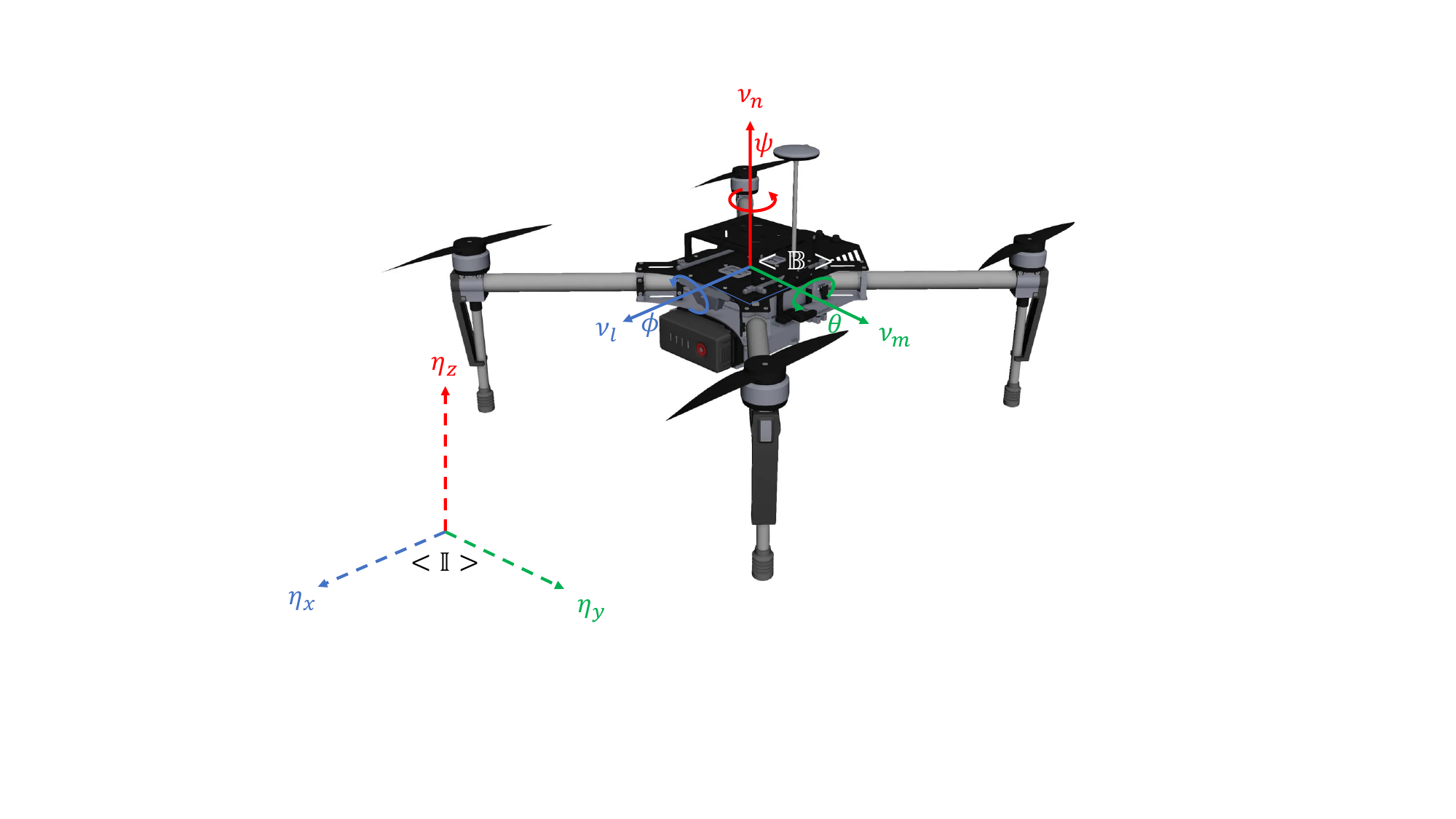}
\caption{UAV reference frame DJI Matrice 100}
\label{Fig:2.1}
\end{figure}

Figure \ref{Fig:2.1} shows the quadcopter platform, where the inertial reference frame, fixed to the world, is labeled as $\langle \mathbb{I} \rangle$ with unit vectors $\left\lbrace \mathbf{I}_x, \mathbf{I}_y, \mathbf{I}_z \right\rbrace$. The body-fixed reference frame, which moves with the quadcopter and is aligned with its motion, is designated as $\langle \mathbb{B} \rangle$ with unit vectors $\left\lbrace \mathbf{B}_x, \mathbf{B}_y, \mathbf{B}_z \right\rbrace$, and its origin is located at the quadcopter's center of mass (CoM).

The dynamic model of the unmanned aerial vehicle (UAV) can be derived using Euler-Lagrange formalism \cite{carelli}, which allows us to describe the equations of motion based on kinetic energy, potential energy, and the generalized forces acting on the system. Specifically, the system's state is represented by $\mathbf{x} = \begin{bmatrix}  \mathbf{q} &  \dot{\mathbf{q}} \end{bmatrix}^\intercal \in \mathbb{R}^{12}$, with $\mathbf{q} = \begin{bmatrix} \bm{\eta}  & \bm{\Omega} \end{bmatrix}^\intercal \in \mathbb{R}^{6}$ and $\dot{\mathbf{q}} = \begin{bmatrix}  \dot{\bm{\eta}}  & \dot{\bm{\Omega}} \end{bmatrix}^\intercal \in \mathbb{R}^{6}$, where $\bm{\eta} = \begin{bmatrix} \eta_x & \eta_y &  \eta_z  \end{bmatrix}^{\intercal}$ denotes the position coordinates, $\dot{\bm{\eta}} = \begin{bmatrix} \dot{\eta}_x & \dot{\eta}_y &  \dot{\eta}_z  \end{bmatrix}$ are the linear velocities in inertial frame, $\bm{\Omega} =\begin{bmatrix} {\phi} & {\Omega} &  {\psi}  \end{bmatrix}^\intercal$ represents the Euler angles (roll, pitch, yaw), and $\dot{\bm{\Omega}} = \begin{bmatrix} \dot{\phi} & \dot{\Omega} &  \dot{\psi}  \end{bmatrix}^\intercal$ are the corresponding angular velocities in body frame.

\subsection{Euler-Lagrange Equations}

To derive the dynamic model, the Lagrangian function must satisfy the Euler-Lagrange equation:

\[
    \frac{d}{dt}\left(\frac{\partial \mathcal{L}}{\partial \dot{\mathbf{q}}} \right)-\frac{\partial \mathcal{L}}{\partial {\mathbf{q}}}=\begin{bmatrix}
       \bm{\mathcal{F}} \\ \bm{\tau}
    \end{bmatrix},
\]

The generalized forces applied in the inertial reference frame are given by $\bm{\mathcal{F}} = \mathbf{R} \mathbf{F} = \begin{bmatrix} ^e f_{x} & ^e f_{y} & ^e f_{z} \end{bmatrix}^\intercal$, while the torques acting on the system in the body reference frame are represented by $\bm{\tau} = \begin{bmatrix} ^b \tau_{\phi} & ^b \tau_{\theta} & ^b \tau_{\psi} \end{bmatrix}^\intercal$ \cite{CASTILLO200741}. Here, $\mathbf{R}$ is the rotation matrix that transforms vectors from the body reference frame to the inertial reference frame, and $\mathbf{F} = \begin{bmatrix} 0 & 0 & ^b f_{z} \end{bmatrix}^\intercal$, where $^b f_{z}$ represents the total thrust along the body's z-axis.

The Lagrangian function, \(\mathcal{L}\), represents the difference between the kinetic and potential energies of the system and is given by:
\begin{equation*}
    \mathcal{L} = \frac{1}{2} m \dot{\bm{\eta}}^\top \dot{\bm{\eta}} + \frac{1}{2} \bm{\omega}^\top \mathbf{I} \bm{\omega} - m g \eta_z,
\end{equation*}

where \(m\) is the mass of the rigid body, \(\dot{\bm{\eta}}\) represents the linear velocity, \(\bm{\omega}\) the angular velocity, and \(\mathbf{I}_\omega\) the inertia tensor. The term \(m g \eta_z\) represents the gravitational potential energy.

Using this framework, the dynamics of a rigid body with six degrees of freedom can be expressed in a compact form as:

\begin{equation}
        \begin{aligned}
    \begin{bmatrix}
        \bm{\mathcal{F}} \\ \bm{\tau}
    \end{bmatrix}
    &= \begin{bmatrix}
        m \mathbf{I}_{3 \times 3} & \mathbf{0}_{3 \times 3} \\
        \mathbf{0}_{3 \times 3} & \mathbf{M}(\bm{\Omega})
    \end{bmatrix} 
    \begin{bmatrix}
        \ddot{\bm{\eta}} \\ \ddot{\bm{\Omega}}
    \end{bmatrix}\\
    &+ \begin{bmatrix}
        \mathbf{0}_{3 \times 3} & \mathbf{0}_{3 \times 3} \\
        \mathbf{0}_{3 \times 3} & \mathbf{C}(\bm{\Omega}, \dot{\bm{\Omega}})
    \end{bmatrix} 
    \begin{bmatrix}
        \dot{\bm{\eta}} \\ \dot{\bm{\Omega}}
    \end{bmatrix} + \begin{bmatrix}
        mg\mathbf{e}_3 \\ 0
    \end{bmatrix},
    \end{aligned}
\end{equation}

Here, $\mathbf{M}(\bm{\Omega})$ and $\mathbf{C}(\bm{\Omega}, \dot{\bm{\Omega}})$ represent the rotational inertia and Coriolis matrices, respectively, and their structures are detailed in the appendix. Additionally, $\mathbf{e}_3$ is a unit vector in the vertical direction.

The dynamic model can be rewritten in its more compact form as:
\begin{equation}\label{Eq: Euler-Lagrange}
    \bar{\mathbf{M}}(\mathbf{q})\ddot{\mathbf{q}} +  \bar{\mathbf{C}}(\mathbf{q},\dot{\mathbf{q}})\dot{\mathbf{q}} + \bar{\mathbf{G}}  = \mathbf{T},
\end{equation}
where \(\bar{\mathbf{M}}(\mathbf{q})\) is the generalized inertia matrix, \(\bar{\mathbf{C}}(\mathbf{q}, \dot{\mathbf{q}})\) is the Coriolis generalized matrix, \(\bar{\mathbf{G}}\) represents the gravitational forces, and $\mathbf{T} = \begin{bmatrix}
    \bm{\mathcal{F}} & \bm\tau \end{bmatrix}^\intercal $ corresponds to the applied forces and torques.

\subsection{PD-Dynamic Identification}

It is assumed that the torques responsible for horizontal motion, \( {^b\bm{\tau}_h} = \begin{bmatrix} \tau_x & \tau_y \end{bmatrix}^\intercal \), are controlled by a low-level attitude PD controller. This controller regulates the torques in the \(x\)- and \(y\)-axes to ensure stability in the horizontal plane. These torques are applied as follows:

\[
^b\bm{\tau}_h = \mathbf{K}_{p_h} (\bm{\Omega}_{h_d} - \bm{\Omega}_h) + \mathbf{K}_{d_h} (\dot{\bm{\Omega}}_{h_d} - \dot{\bm{\Omega}}_h),
\]
where \(\bm{\Omega}_h = \begin{bmatrix} \phi & \theta \end{bmatrix}^\top\) represents the current roll and pitch angles, and \(\bm{\Omega}_{h_d} = \begin{bmatrix} \phi_d & \theta_d \end{bmatrix}^\top\) represents the desired attitude. Assuming no change in the desired attitude rate, \(\dot{\bm{\Omega}}_{h_d} = 0\). The proportional and derivative gains are denoted by \(\mathbf{K}_{p_h}\) and \(\mathbf{K}_{d_h}\), respectively.

It is assumed that if \(\bm{\Omega}_{h_d} = 0\), then:

\[
\lim_{t \to \infty} \| \bm{\tau}_h \| = 0 \quad \text{and} \quad \lim_{t \to \infty} \| \Tilde{\bm{\Omega}}_h \| = 0,
\]
where \(\Tilde{\bm{\Omega}}_h = \bm{\Omega}_{h_d} - \bm{\Omega}_h\). This ensures that over time, the applied torque and attitude error in the horizontal plane will tend to zero.

For yaw control, the torque is governed by a PD controller with acceleration feedback:

\[
^b\tau_\psi = k_{p_\psi} (\dot{\psi}_d - \dot{\psi}) + k_{d_\psi} (\ddot{\psi}_d - \ddot{\psi}),
\]
where \(\dot{\psi}_d\) is the desired yaw rate, \(\ddot{\psi}_d\) is the desired yaw acceleration, \(\dot{\psi}\) is the current yaw rate, and \(\ddot{\psi}\) is the current yaw acceleration. The yaw control considers both yaw rate and yaw acceleration. Therefore, we assume that:

\[
\lim_{t \to \infty} \| \tau_\psi \| = 0 \quad \text{and} \quad \lim_{t \to \infty} \| \Tilde{\psi} \| = 0,
\]
where \(\Tilde{\psi} = \psi_d - \psi\), implying that both the applied yaw torque and yaw angle error will tend to zero over time.

In hover, with gravity compensation, the vertical thrust force is given by:

\[
{^b f_z} = k_{p_z} (\dot{\eta}_{z_d} - \dot{\eta}_z) + k_{d_z} (\ddot{\eta}_{z_d} - \ddot{\eta}_z) + mg,
\]
where \(\dot{\eta}_{z_d}\) is the reference vertical velocity, \(\ddot{\eta}_{z_d}\) and \(\ddot{\eta}_z\) are the time derivatives of the reference and actual vertical velocities, and \(k_{p_z}\) and \(k_{d_z}\) are the proportional and derivative gains. It is assumed that if \(\dot{\eta}_{z_d} = 0\), then:

\[
\lim_{t \to \infty} \| {^b f_z} \| = mg \quad \text{and} \quad \lim_{t \to \infty} \| \dot{\Tilde{\eta}}_z \| = 0,
\]
where \(\dot{\Tilde{\eta}}_z = \dot{\eta}_{z_d} - \dot{\eta}_z\), implying that the thrust force stabilizes at the hover value and the vertical velocity error tends to zero.

Based on the above, the general PD controller structure with vertical velocity and attitude angles as inputs can be expressed as follows:
\begin{equation}\label{eq:PD_complete}
    \begin{bmatrix}
^ef_{{x}} \\ ^ef_{{y}} \\ ^ef_{{z}}\\ ^b\tau_{{\phi}} \\ ^b\tau_{\theta} \\ ^b\tau_{\psi}   
\end{bmatrix} = \begin{bmatrix}
     \mathbf{R}\begin{bmatrix}
        0 \\ 0 \\ k_{p_z}(\dot\eta_{z_{d}} - \dot\eta_z) + k_{d_z}(\ddot\eta_{z_{d}} - \ddot\eta_z) + m g
    \end{bmatrix}  \\ k_{p_\phi}({\phi}_{d} - \phi) + k_{p_\phi} (\dot{\phi}_{d} - \dot\phi) \\ k_{p_\theta}({\theta}_{d} - \theta) + k_{p_\theta} (\dot{\theta}_{d} - \dot\theta) \\ k_{p_\psi}(\dot{\psi}_{d} - \dot\psi) + k_{p_\psi} (\ddot{\psi}_{d} - \ddot\psi)
\end{bmatrix}
\end{equation}

where \(\mathbf{R}\) is the rotation matrix between the body and the inertial frame.

The equation (\ref{eq:PD_complete}) can be rewritten as follows:

\begin{equation}\label{Eq: PD}
     \mathbf{T}_{PD} = \bar{\mathbf{R}} \left[\bar{\mathbf{S}}_1 \bm\mu - \bar{\mathbf{S}}_2 \mathbf{q} - \bar{\mathbf{S}}_3 \dot{\mathbf{q}} - \bar{\mathbf{S}}_4 \ddot{\mathbf{q}} + \bar{\mathbf{S}}_5  \right]
\end{equation}

Given the limited access to the unmanned aerial vehicle (UAV), specifically the DJI Matrice 100, the state vector is formulated based on the available signals. The input signals applied to the quadcopter are represented by \( \mathbf{T}_{PD} := \begin{bmatrix} ^e\bm{\mathcal{F}} & ^b\bm\tau \end{bmatrix}^\intercal \in \mathbb{R}^{6} \) and \( \bm\mu = \begin{bmatrix} 0 & 0 & \dot\eta_{z_d} & \phi_{d} & \theta_{d} & \dot\psi_{d} \end{bmatrix}^\intercal \), which correspond to the signals received from the UAV system using the Onboard SDK based on ROS. The matrices \(\bar{\mathbf{S}}_1 = \text{diag}( 0,0, \xi_1, \xi_2, \xi_3, \xi_4 )\), \(\bar{\mathbf{S}}_2 = \text{diag}( 0,0, \xi_5, \xi_6 )\), \(\bar{\mathbf{S}}_3 = \text{diag}( 0,0, \xi_7, \xi_8, \xi_9, \xi_{10} )\), \(\bar{\mathbf{S}}_4 = \text{diag}( 0,0, \xi_{11}, 0, \xi_{12} )\), and \(\bar{\mathbf{S}}_5 = \begin{bmatrix} 0 & 0 & \xi_{13} & 0 & 0 & 0 \end{bmatrix}^\intercal\) are diagonal matrices that encapsulate the relevant parameters. The inertias \(I_x\), \(I_y\), \(I_z\) are included as parameters \(\xi_{14}-\xi_{16}\), which must also be identified.  

Moreover, the matrix \( \bar{\mathbf{R}} \) is an expanded rotation matrix, expressed as:
\[
\bar{\mathbf{R}} = \begin{bmatrix}
    \mathbf{R} & \mathbf{0}_{3\times3} \\
    \mathbf{0}_{3\times3} & \mathbf{0}_{3\times3}
\end{bmatrix}.
\]

This formulation reflects the limited information available on the UAV and how these signals are incorporated to describe the dynamic system accurately.

If dynamic model in (\ref{Eq: Euler-Lagrange}) is equated to (\ref{Eq: PD}), we obtain:
\[
\bar{\mathbf{M}}\ddot{\mathbf{q}} +  \bar{\mathbf{C}}\dot{\mathbf{q}} + \bar{\mathbf{G}} = \bar{\mathbf{R}} \left[\bar{\mathbf{S}}_1 \bm\mu - \bar{\mathbf{S}}_2 \mathbf{q} - \bar{\mathbf{S}}_3 \dot{\mathbf{q}} - \bar{\mathbf{S}}_4 \ddot{\mathbf{q}} + \bar{\mathbf{S}}_5  \right]
\]

where the dynamic model based on vertical velocity and desired attitude is expressed as follows:
\begin{equation}
\resizebox{\linewidth}{!}{$
\ddot{\mathbf{q}} = \left( \bar{\mathbf{M}} +  \bar{\mathbf{R}}\mathbf{S}_4\right)^{-1} \left[ \bar{\mathbf{R}} \left(\bar{\mathbf{S}}_1 \bm\mu - \bar{\mathbf{S}}_2 \mathbf{q} - \bar{\mathbf{S}}_3 \dot{\mathbf{q}} + \bar{\mathbf{S}}_5 \right) -    \bar{\mathbf{C}}\dot{\mathbf{q}} - \bar{\mathbf{G}} \right]
$}
\end{equation}

\section{Sparse Identification of Nonlinear Dynamics}

The \textit{Sparse Identification of Nonlinear Dynamics (SINDy)} method provides a data-driven framework for identifying parsimonious models of dynamical systems \cite{Brunton2019}. The core assumption of SINDy is that most dynamical systems can be represented by only a few active terms, while the majority of potential candidate functions remain inactive. This approach leverages sparse regression techniques to select the most relevant terms in the governing equations of the system.

\subsection{Mathematical Formulation}

Consider a nonlinear dynamical system governed by the ordinary differential equation (ODE):
\[
\dot{\mathbf{x}}(t) = \mathbf{f}(\mathbf{x}(t), \mathbf{u}(t)),
\]

where \(\mathbf{x}(t) \in \mathbb{R}^n\) is the state vector, \(\mathbf{u}(t) \in \mathbb{R}^q\) is the control input, and \(\mathbf{f}: \mathbb{R}^n \to \mathbb{R}^n\) represents the nonlinear dynamics. The objective of SINDy is to find a sparse approximation of \(\mathbf{f}(\mathbf{x}, \mathbf{u})\) using time-series data of \(\mathbf{x}(t)\) and its derivatives \(\dot{\mathbf{x}}(t)\).

The goal is to express \(\mathbf{f}(\mathbf{x},\mathbf{u})\) as a sparse linear combination of the columns of \(\Theta(\mathbf{x},\mathbf{u})\). This leads to the following approximation:
\begin{equation}
    \dot{\mathbf{X}} \approx \Theta(\mathbf{X}, \mathbf{U}) \mathbf{\Xi},
\end{equation}

where \(\dot{\mathbf{X}} \in \mathbb{R}^{m \times n}\) contains the derivative data, \(\Theta(\mathbf{X}, \mathbf{U}) \in \mathbb{R}^{m \times p}\) is the evaluation of the function library over the dataset where \(p\) is the number of candidate functions; \(\mathbf{U} \in \mathbb{R}^{m \times q}\) contains the control input data, and \(\mathbf{\Xi} \in \mathbb{R}^{p \times n}\) is the sparse matrix that includes both the dynamics of the system and the effects of the control inputs. 

By including control terms in the function library, SINDy can identify the influence of external forces or control actions on the system dynamics, making it suitable for applications in robotics, autonomous systems, and control theory.

The library of candidate functions can include polynomial, trigonometric, and other nonlinear functions of the states \(\mathbf{x}(t)\) and can be expressed as:
\[
\Theta = \left[ 1 \  \mathbf{x} \  \mathbf{u} \  \mathbf{x}^2 \  \mathbf{x} \mathbf{u} \  \sin(\mathbf{\Theta}) \  \cos(\mathbf{\Theta}) \  \mathbf{x} \sin(\mathbf{\Theta}) \  \mathbf{x} \cos(\mathbf{\Theta}) \right]
\]

To determine \(\mathbf{\Xi}\), a sparse regression problem is formulated as:
\[
\min_{\mathbf{\Xi}} \|\dot{\mathbf{X}} - \Theta(\mathbf{X}, \mathbf{U}) \mathbf{\Xi}\|_2^2 + \lambda \|\mathbf{\Xi}\|_1,
\]

Thus, the full identified dynamics with control are:
\begin{equation}
    \begin{split}
        \dot{\mathbf{x}} = 
\underbrace{\Xi_1}_{\text{Constant}} + 
\underbrace{\Xi_2 \mathbf{x}}_{\text{Linear}} +
\underbrace{\Xi_3 \mathbf{u}}_{\text{Control}} +
\underbrace{\Xi_4 \mathbf{x}^2}_{\text{Quadratic}} +  
\underbrace{\Xi_6 \mathbf{x} \mathbf{u}}_{\text{Cross Terms}}  + \\ \underbrace{\Xi_7 \sin(\mathbf{\Theta}) + \Xi_8 \cos(\mathbf{\Theta})}_{\text{Trigonometric}} + 
\underbrace{\Xi_9 \mathbf{x} \sin(\mathbf{\Theta}) + \Xi_{10} \mathbf{x} \cos(\mathbf{\Theta})}_{\text{Mixed}}.
    \end{split}
\end{equation}
where the \(\Xi_i\) terms represent constant, linear, quadratic, cross, and trigonometric interactions between the state vector, control inputs, and Euler angles, capturing the essential dynamics and control effects.
In practical applications, the time derivatives \(\dot{\mathbf{x}}(t)\) are not directly measured and must be calculated through numerical differentiation of the state data.

\section{System Identification}

\begin{figure*}[ht]
\centering
\begin{minipage}{\linewidth}
    \centering
    \includegraphics[clip, trim=0cm 0cm 0cm 0cm, width=\linewidth]{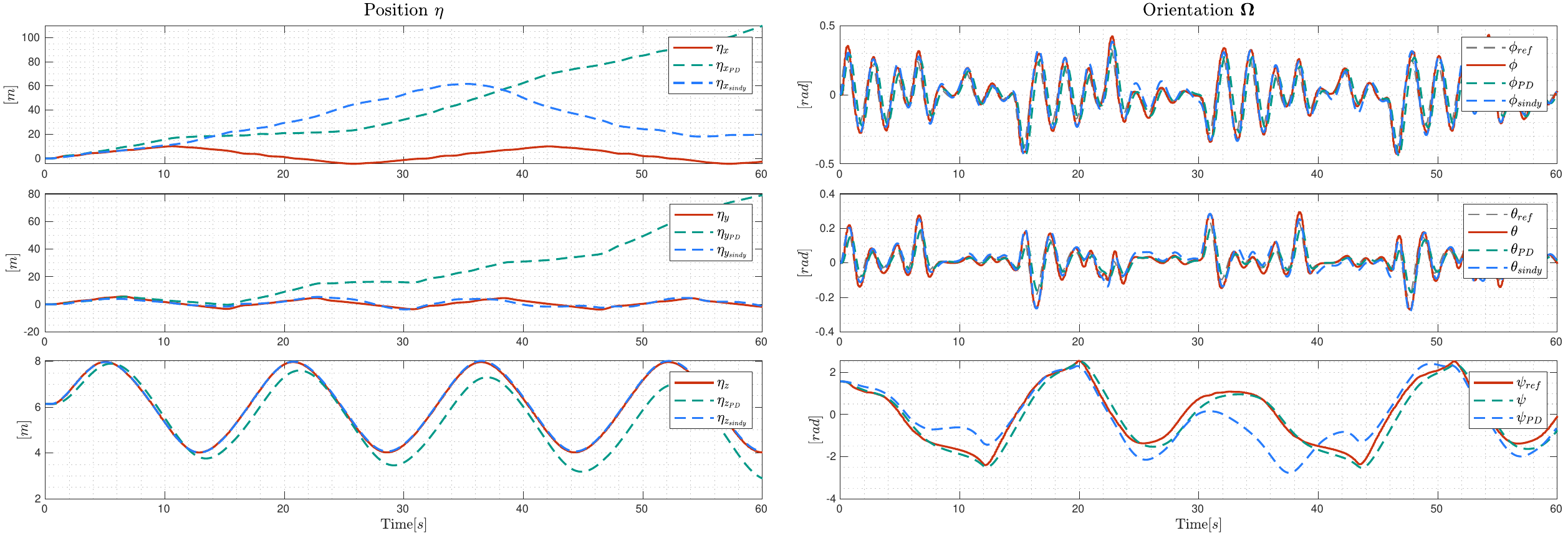}
    \caption{Comparison of the UAV's linear and angular positions with respect to the reference values.}
    \label{Fig:pos_ref}
    \vspace{0.5cm}
\end{minipage}
\begin{minipage}{\linewidth}
    \centering
    \includegraphics[clip, trim=0cm 0cm 0cm 0cm, width=\linewidth]{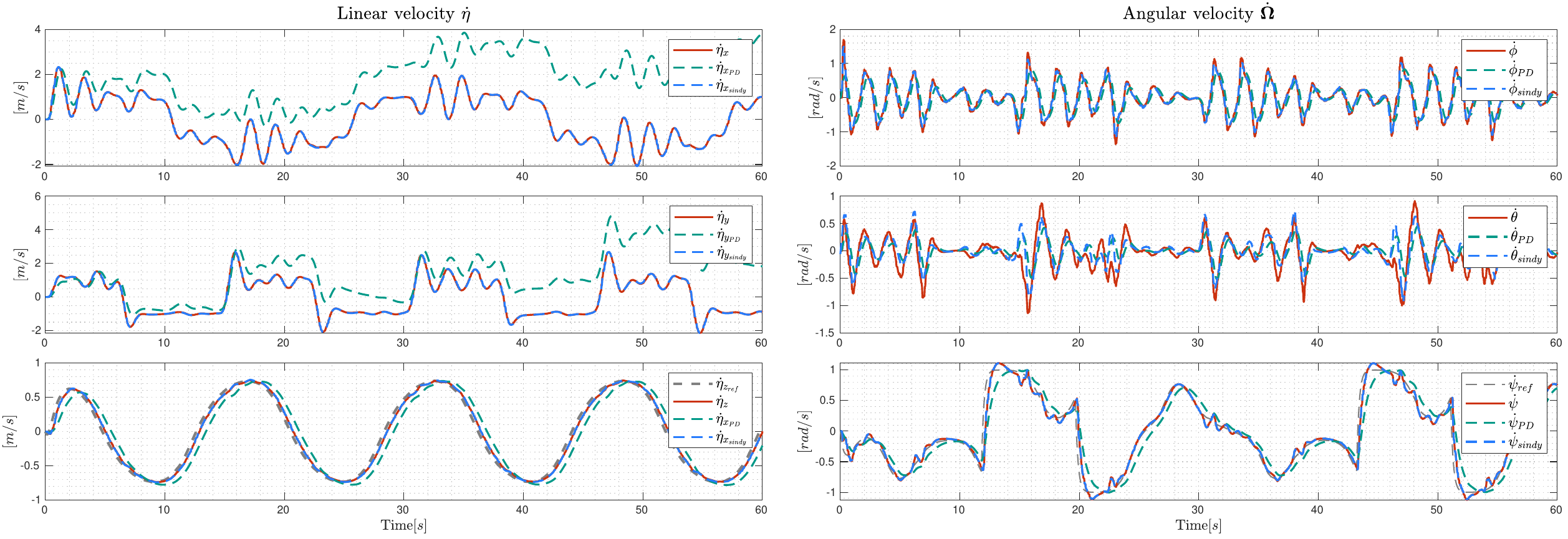}
    \caption{Comparison of the UAV's linear and angular velocities with respect to the reference values.}
    \label{Fig:vel_ref}
\end{minipage}
\end{figure*}

The flight control system of a multicopter is nonlinear and underactuated, with significant coupling between its degrees of freedom. This dynamic behavior can be effectively approximated using a PD control structure. Forces and torques derived from the Euler-Lagrange method are transformed into control inputs through an optimization-based identification process. Additionally, the nonlinear dynamics are further identified using the SINDy method, which allows for a data-driven discovery of the governing equations, enhancing the predictive capability of the control system.

The state vector \(\mathbf{x}(t)\) is obtained from the velocity and acceleration measurements collected from the UAV during experimental flights from a DJI MATRICE 100 quadcopter, which features an integrated low-level flight controller.

\begin{figure}[H]
\centering
\includegraphics[clip, trim=11.0cm 3cm 12.6cm 5.1cm, width=5cm]{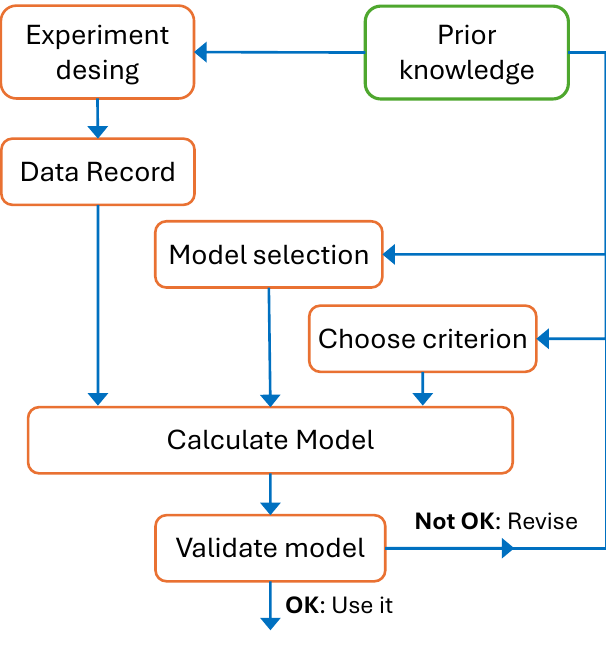}
\caption{System identification process for the UAV.}
\label{Fig:ident_proceso}
\end{figure}

This approach uses experimental data from real-world tests to system identification, as depicted in Fig~\ref{Fig:ident_proceso}, involves key steps that guide the development of mathematical models from observed data \cite{Quan2017}. 

\subsection{Euler-Lagrange PD approximation}
\begin{table*}[ht]
\centering
\caption{Comparison of RMSE, MAE, and Concordance Index for SINDy and PD}\label{tab: Comparison}
\begin{tabular*}{\textwidth}{@{\extracolsep{\fill}}lcccccc}
\toprule
& \multicolumn{2}{@{}c@{}}{\textbf{RMSE}} & \multicolumn{2}{@{}c@{}}{\textbf{MAE}} & \multicolumn{2}{@{}c@{}}{\textbf{Concordance Index}} \\
\cmidrule{2-3} \cmidrule{4-5} \cmidrule{6-7}
State & \textbf{SINDy} & \textbf{PD} & \textbf{SINDy} & \textbf{PD} & \textbf{SINDy} & \textbf{PD} \\
\midrule
$\eta_x$         & 33.1389  & 53.4461   & 27.1978  & 41.6780   & 0.1300  & 0.0887  \\
$\eta_y$         & 1.2704   & 32.8757   & 1.0249   & 23.7101   & 0.7435  & 0.0779  \\
$\eta_z$         & 0.0399   & 0.6811    & 0.0334   & 0.5703    & 0.9861  & 0.7705  \\
$\dot{\eta}_x$   & 0.0179   & 1.0154    & 0.0133   & 0.9151    & 0.9927  & 0.1133  \\
$\dot{\eta}_y$   & 0.0252   & 1.1049    & 0.0172   & 1.0193    & 0.9916  & 0.0515  \\
$\dot{\eta}_z$   & 0.0042   & 1.1453    & 0.0035   & 0.9436    & 0.9964  & 0.4652  \\
$\phi$           & 0.0290   & 2.1226    & 0.0231   & 1.8431    & 0.9075  & 0.0575  \\
$\theta$         & 0.0238   & 1.9420    & 0.0194   & 1.5267    & 0.8333  & 0.0355  \\
$\psi$           & 0.8033   & 0.9464    & 0.6057   & 0.7830    & 0.7498  & 0.5414  \\
$\dot{\phi}$     & 0.0438   & 0.2639    & 0.0285   & 0.1991    & 0.9597  & 0.6997  \\
$\dot{\theta}$   & 0.1788   & 0.2468    & 0.1224   & 0.1714    & 0.6744  & 0.4701  \\
$\dot{\psi}$     & 0.0049   & 0.1684    & 0.0033   & 0.1343    & 0.9966  & 0.8588  \\
\midrule
$\mathbf{x}$  & \textbf{9.5764} & \textbf{18.1445} & \textbf{2.4244} & \textbf{6.1245} & \textbf{0.8301} & \textbf{0.3525} \\
\bottomrule
\end{tabular*}
\end{table*}

The objective is to minimize the quadratic error between the forces and torques generated by the nonlinear model based on the Euler-Lagrange method and those generated by the parametrized model, which is based on the dynamic parameters and PD constants $\bm{\xi} \in \mathbb{R}^{16}$, encapsulating the UAV's dynamic properties.
 Quadratic programming techniques are applied iteratively to adjust the parameters of the diagonal matrices \(\bar{\mathbf{S}}_1, \bar{\mathbf{S}}_2, \bar{\mathbf{S}}_3, \bar{\mathbf{S}}_4, \bar{\mathbf{S}}_5\) until the cost function \( J \) reaches a minimum. The optimization problem is formulated as follows:
\[
\min_{\bm\xi} \int_0^{t+t_f} J = \left\|\mathbf{T}(\bm\xi) - \mathbf{T}_{\text{PD}}(\bm\xi)\right\|^2 dt,
\]

The presence of noise in the measurements is addressed using a low-pass filter of the form \( \frac{\lambda}{s + \lambda} \), with \(\lambda = 1\), applied to both the system's measurements and the reference values. This ensures that the noise does not affect the identification process, and since the time derivatives \(\dot{x}(t)\) are not directly measured, they must be estimated from the state data and pass through the same filter.

The identified diagonal matrices are presented in Table~\ref{tab:xi_table}:
\begin{table}[h]
\centering
\caption{Values of $\xi$}\label{tab:xi_table}%
\begin{tabular}{@{}llll@{}}
\toprule
$\xi_1=0.6756$  & $\xi_2=1.0000$  & $\xi_3=0.6344$  & $\xi_4=1.0000$  \\
$\xi_5=0.4080$  & $\xi_6=1.0000$  & $\xi_7=1.0000$  & $\xi_8=1.0000$  \\
$\xi_9=0.2953$  & $\xi_{10}=0.5941$ & $\xi_{11}=-0.8109$ & $\xi_{12}=1.0000$  \\
$\xi_{13}=0.3984$  & $\xi_{14}=0.00546$ & $\xi_{15}=0.0035$ & $\xi_{16}=0.00659$ \\
\bottomrule
\end{tabular}
\end{table}

These values provide the best approximation of the UAV's dynamic behavior and are suitable for further modeling and MPC design. The system's dynamics are characterized in a concise form based on the control inputs, and the identified model is formulated as a vector function, given by:
\begin{equation}
    \dot{\mathbf{x}} = \mathbf{f}({\mathbf{x}}, \bm\mu, \bm\xi).
\end{equation}

\subsection{SINDy Approximation}
Given the dataset of states \(\mathbf{x}(t)\), control inputs \(\mathbf{u}(t)\), and numerically computed derivatives \(\dot{\mathbf{x}}(t)\), the SINDy algorithm outputs a sparse coefficient matrix \(\mathbf{\Xi}\), which contains mostly zeros, with nonzero entries corresponding to the active terms that describe the system’s true dynamics. This matrix identifies the relevant terms in the function library \(\Theta(\mathbf{x}, \mathbf{u})\) that contribute to the dynamics. This work employed numerical differentiation techniques and smoothing methods, such as the Savitzky-Golay filter, to obtain \(\dot{\mathbf{x}}(t)\), as facilitated by the approach in \cite{Kaheman2020}.

The problem is solved using the SR3 optimization method, which applies both \(l_1\) and \(l_2\) regularization to enforce sparsity. SR3 aligns well with the desired objective, as it promotes a parsimonious model by eliminating insignificant terms \cite{Kaheman2020, Zheng2019}. While alternative methods like STLSQ use \(l_2\) regularization (Ridge), SR3 ensures better sparsity by introducing a thresholding mechanism and regularization gain \(\lambda = 0.05\), resulting in a compact model for \(\mathbf{f}(\mathbf{x}, \mathbf{u})\).

Fig. \ref{Fig:pos_ref} and Fig. \ref{Fig:vel_ref} shows the results of the identification process, where the suffix \textit{ref} refers to the reference values received by the UAV, the suffix \textit{sindy} refers to the values generated by the SINDy-identified model, and the suffix \textit{PD} corresponds to the Euler-Lagrange approximation model. Table \ref{tab: Comparison} presents a comparison of the RMSE, MAE, and Concordance Index for both the SINDy and PD models. The comparison is performed specifically for each state, as well as globally for the full state vector.

\subsection{Validation using MPC}

 The validation of the identified models (PD-based and SINDy) is performed in a real-world experimental setup, where the UAV is required to follow a predefined reference trajectory, defined in Table \ref{tab: trajectory}:
\begin{table}[H]
\centering
\caption{Desired positions for different trajectory types}
\begin{tabular}{|c|c|c|c|}
\hline
\textbf{Trajectory} & $x_d(t)$ & $y_d(t)$ & $z_d(t)$ \\ \hline
\textbf{Sinusoidal} & $4 \sin(0.32 t) + 3$ & $4 \sin(0.64 t)$ & $2 \sin(0.64 t) + 6$ \\ 
\textbf{Circular}   & $5 \cos(t)$ & $5 \sin(t)$ & 6 \\ 
\textbf{Spiral} & $e^{0.06 t} \cos(t)$ & $e^{0.06 t} \sin(t)$ & $0.1 t + 5$ \\ \hline
\end{tabular}
\label{tab: trajectory}
\end{table}
Trajectory tracking is achieved using MPC, which optimizes the control inputs over a finite prediction horizon to minimize tracking errors and control effort. The cost function for OCP is defined as:
\begin{subequations}
\begin{align} 
\min_{\Tilde{\mathbf{x}}(.)\, {\bm\mu}(.)} & \quad \frac{1}{2} \Tilde{\mathbf{x}}^{\intercal}_N \mathbf{Q} \Tilde{\mathbf{x}}_N   + \sum_{k=0}^{N-1}   \frac{1}{2} \Tilde{\mathbf{x}}^{\intercal}_k \mathbf{Q} \Tilde{\mathbf{x}}_k + \frac{1}{2}{\bm\mu}^{\intercal}_k \mathbf{R} {\bm\mu}_k \label{eq:2.43a} \\
\textrm{s.t.: } & \quad {\bm{x}}_{k+1} = \mathbf{f}_{din}( \bm{x}_k , \bm\mu_k), \quad \forall{k} \in [0, N-1] \label{eq:2.43c} \\
& \quad \bm x_{0} = [\bm \eta_0~ \bm \Omega_0 ]^\intercal  \label{eq:2.43c} \\
& \quad |\dot\eta_{z_d}| \leq 5 \, \text{m/s} \\
& \quad |\phi_d| \leq 0.611 \, \text{rad}, \\
& \quad |\theta_d| \leq 0.611 \, \text{rad} \\
& \quad |\dot\psi_d| \leq \frac{5}{6} \pi \, \text{rad/s}
\end{align}
\end{subequations}

Here, \( \mathbf{x}_k \) represents the state vector at time step \( k \), and \( \Tilde{\mathbf{x}} = \mathbf{x}_k - \mathbf{x}_{\text{ref,k}} \) denotes the deviation from the reference state vector \( \mathbf{x}_{\text{ref,k}} \). The term \( \mathbf{Q} = 2\mathbf{I}_{3\times3} \) is the state weighting matrix, and \( \mathbf{R} = 0.1\mathbf{I}_{3\times3}\) is the control input weighting matrix. The prediction horizon is \( N \), and the system dynamics are modeled by \( \mathbf{f}_{din}(\bm{x}_k, \bm\mu_k) \). The initial conditions for the system are given by $\mathbf{x}_0 = \begin{bmatrix}
3&3&5&0&0&0 \end{bmatrix}^\intercal $. The control input constraints, such as the limits on vertical velocity, desired angles, and yaw rate, are imposed due to the limitations of the DJI Onboard SDK, which governs the control signals received by the UAV.

The optimization is solved at each time step, and the first control input is applied in a receding horizon of \(N = 30\), with a sampling frequency of 30 Hz. The optimization is performed using Acados and CasADi through a sequential quadratic programming (SQP) method, executed within a real-time iteration (RTI) scheme. The system communicates using ROS (Robot Operating System), enabling real-time communication within the UAV architecture. The onboard PC of the UAV is equipped with a Jetson Orin 16 GB (275 TOPS), ensuring accurate real-time control.

\subsection{Analysis of Results}

\begin{figure*}[t]
\centering
\includegraphics[clip, trim=0cm 0cm 0cm 0cm, width=\linewidth]{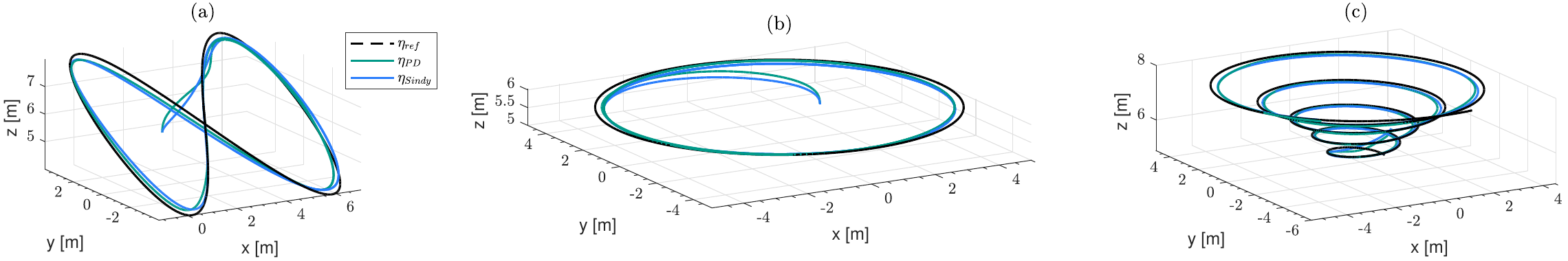}
\caption{Spatial behavior of the UAV under the PD-Euler Lagrange and SINDy models for three different reference trajectories: (a) Sinusoidal, (b) Circular, and (c) Spiral. The black dashed line represents the desired reference trajectory, while the blue and green lines correspond to the actual trajectories tracked.}
\label{Fig:Comportamiento}
\end{figure*}

\begin{figure*}[t]
\centering
\includegraphics[clip, trim=0cm 0cm 0cm 0cm, width=\linewidth]{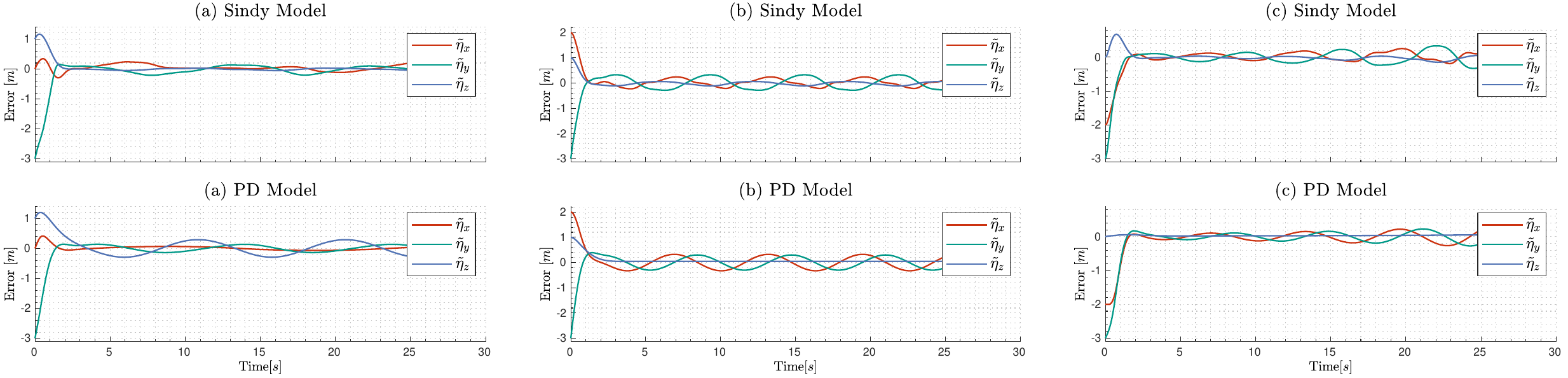}
\caption{Control errors in (a) Sinusoidal (b) Circular and (c) Spiral trayectories.}
\label{Fig: Error control}
\end{figure*}

The results of the validation process were analyzed by examining the control errors and the corresponding control actions during the trajectory tracking task. The following aspects were evaluated:

\subsubsection{Overall Performance}

Figure~\ref{Fig:Comportamiento} presents the spatial behavior during the execution of the desired task for both models in the trajectory tracking task. These results highlight the effectiveness of the identified models when integrated into an MPC framework for real-world UAV control applications.

The real-time implementation of the MPC controller, combined with the data-driven models, enabled precise UAV control with minimal tracking error and reduced control effort. This highlights the potential of the identified models to improve performance in dynamic environments.

\subsubsection{Control Errors}

The position errors across the three experimental setups can be observed in Fig. \ref{Fig: Error control}. The errors demonstrated that both the PD-based and SINDy models were able to accurately track the reference trajectory.

Additionally, Table \ref{tab:comparison_error_experiments} presents the errors obtained in tracking the desired trajectory during the MPC as part of the validation process. It provides a comparative analysis of RMSE, MAE, and Concordance Index for both models, where SINDy consistently demonstrates superior performance in terms of tracking accuracy. SINDy exhibits lower error metrics, although the Concordance Index remains high for both models, indicating strong alignment with the reference trajectory. However, SINDy shows a marginally better consistency across all experiments. These results suggest that SINDy offers a more accurate and reliable solution, making it the preferred model in this validation process.
\begin{table}[H]
\centering
\caption{Comparative Results for MPC validation}
\begin{tabular}{|c|c|c|c|c|}
\hline
\textbf{Experiment} & \textbf{Model} & \textbf{RMSE} & \textbf{MAE} & \textbf{Concordance Index} \\
\hline

\multirow{2}{*}{\textbf{Sinusoidal}} & \textbf{PD}    & 0.4823       & 0.4319       & 0.9868 \\
                                      & \textbf{SINDy} & 0.4791       & 0.3372       & 0.9869 \\
\hline
\multirow{2}{*}{\textbf{Circular}}   & \textbf{PD}    & 0.5209       & 0.5560       & 0.9895 \\
                                      & \textbf{SINDy} & 0.4842       & 0.4752       & 0.9909 \\
\hline
\multirow{2}{*}{\textbf{Spiral}}     & \textbf{PD}    & 0.5711       & 0.4068       & 0.9704 \\
                                      & \textbf{SINDy} & 0.4958       & 0.4239       & 0.9777 \\
\hline
\multirow{2}{*}{\textbf{Average}}                     & \textbf{PD}    & 0.5248       & 0.4649       & 0.9822 \\
                                      & \textbf{SINDy} & 0.4864       & 0.4121       & 0.9852 \\
\hline
\end{tabular}
\label{tab:comparison_error_experiments}
\end{table}

\subsubsection{Time Loop}

\begin{figure}[H]
\centering
\includegraphics[clip, trim=0cm 0cm 0cm 0cm, width=1\linewidth]{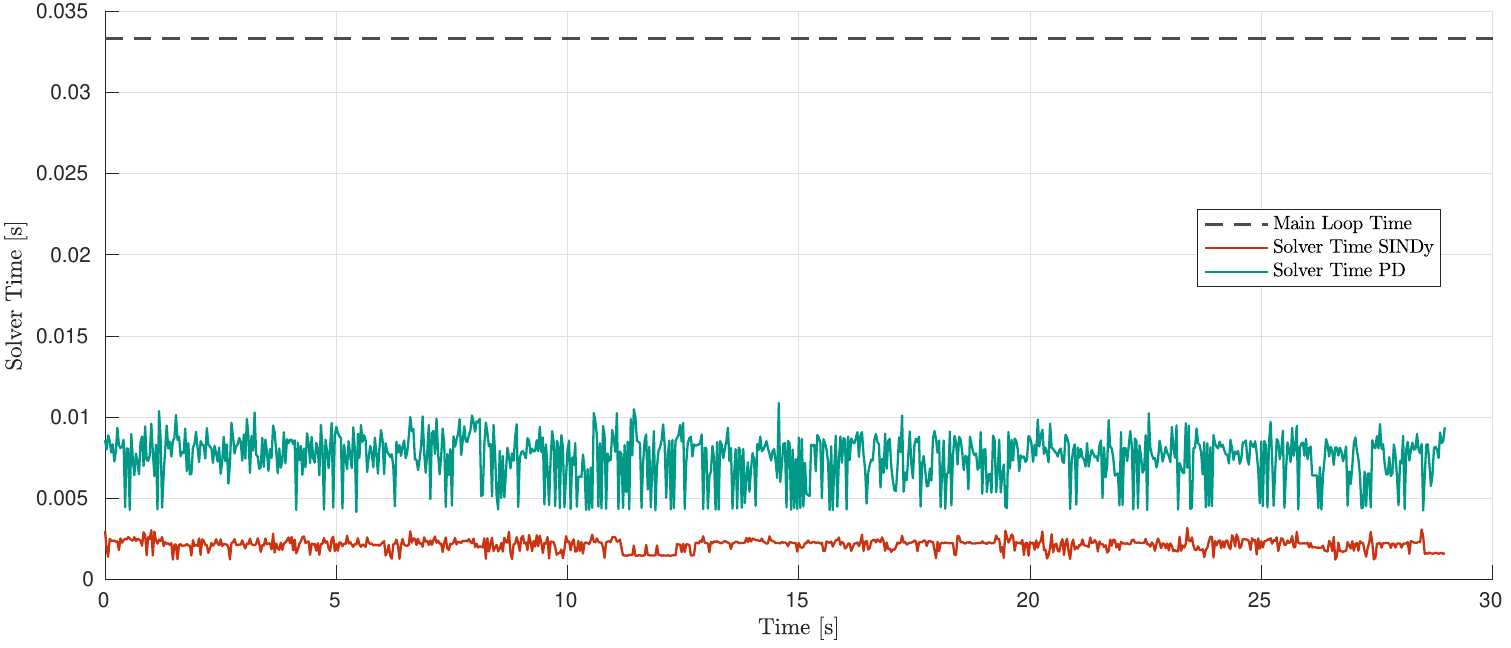}
\caption{Time loop comparative.}
\label{Fig:Time_loop}
\end{figure}

It is noteworthy that the SINDy model converges faster than the PD-Euler Lagrange approximation, as shown in Fig. \ref{Fig:Time_loop}. This is because the latter involves a more complex and nonlinear structure compared to the simpler formulation used by the SINDy model. The reduced computational complexity of the SINDy model enables faster execution, making it more efficient for real-time applications, despite a trade-off in steady-state error performance.

\section{Conclusion}

The findings indicate that the SINDy-based model consistently demonstrates superior performance compared to the PD-based model across different experiments. While both models show strong tracking capabilities, the SINDy model achieves lower error metrics, including RMSE and MAE, highlighting its advantage in trajectory tracking accuracy. Furthermore, the Concordance Index remains high for both models, indicating their robustness in maintaining alignment with the reference trajectory. However, SINDy presents better consistency, as evidenced by its lower error metrics across all experiments, suggesting it is more reliable in varied operating conditions.

In addition to its accuracy, the SINDy model provides computational benefits. Its simpler structure allows for reduced computational complexity, making it better suited for real-time applications where processing speed is critical. On the other hand, the PD model, though useful, has a more complex structure that can slow down computations. Therefore, the SINDy model not only offers improved precision but also ensures faster execution, making it preferable for applications requiring both accuracy and efficiency.

Finally, it is important to note that while the SINDy model performs well under the tested conditions, its accuracy could degrade if the UAV operates outside the parameter space used for model identification. Maintaining consistency between the identification and operating conditions is essential to ensure reliable performance in real-world applications.

\section*{Acknowledgments}
The authors are grateful for the support of INAUT - UNSJ, which provided them with the facilities and workspace necessary to carry out this research.

\appendices
\section{Detailed Representation of Inertia and Coriolis Matrices}

The matrix $\mathbf{M}(\bm{\Omega})$ and its components are given by:

\begin{equation}
\resizebox{\linewidth}{!}{$
\mathbf{M}(\bm{\Omega}) = 
\begin{bmatrix}
    I_{x} & 0 & -I_{x}S_{\theta} \\
    0 & I_{z} + (I_{y} - I_{z})C_{\phi}^2 & C_{\phi}S_{\phi}C_{\theta}(I_{y} - I_{z}) \\
    -I_{x}S_{\theta} & C_{\phi}S_{\phi}C_{\theta}(I_{y} - I_{z}) & I_{z}C_{\phi}^2C_{\theta}^2 + I_{y}S_{\phi}^2C_{\theta}^2 + I_{x}S_{\theta}^2
\end{bmatrix}
$}
\end{equation}

The Coriolis matrix \(\mathbf{C}(\bm{\Omega}, \dot{\bm{\Omega}})\) contains the gyroscopic and centripetal terms and is given by:

\[
\mathbf{C}(\bm{\Theta}, \dot{\bm{\Theta}}) = 
\begin{bmatrix}
    c_{11} & c_{12} & c_{13}\\
    c_{21} & c_{22} & c_{23}\\
    c_{31} & c_{32} & c_{33}
\end{bmatrix},
\]
where the elements of the Coriolis matrix are defined as:

\[
\begin{aligned}
    c_{11} &= 0, \\
    c_{12} &= (I_{y} - I_{z})(\theta_p C_{\phi} S_{\phi} + \psi_p S_{\phi}^2 C_{\theta}) \\
           &\quad + (I_{z} - I_{y})(\psi_p C_{\phi}^2 C_{\theta}) - I_{x} \psi_p C_{\theta}, \\
    c_{13} &= (I_{z} - I_{y}) \psi_p C_{\phi} S_{\phi} C_{\theta}^2, \\
    c_{21} &= (I_{z} - I_{y})(\theta_p C_{\phi} S_{\phi} + \psi_p S_{\phi} C_{\theta}) \\
           &\quad + (I_{y} - I_{z})(\psi_p C_{\phi}^2 C_{\theta}) + I_{x} \psi_p C_{\theta}, \\
    c_{22} &= (I_{z} - I_{y}) \phi_p C_{\phi} S_{\phi}, \\
    c_{23} &= -I_{x} \psi_p S_{\theta} C_{\theta} + I_{y} \psi_p S_{\phi}^2 S_{\theta} C_{\theta} \\
           &\quad + I_{z} \psi_p C_{\phi}^2 S_{\theta} C_{\theta}, \\
    c_{31} &= (I_{y} - I_{z}) \psi_p C_{\theta}^2 S_{\phi} C_{\phi} - I_{x} \theta_p C_{\theta}, \\
    c_{32} &= (I_{z} - I_{y})(\theta_p C_{\phi} S_{\phi} S_{\theta} + \phi_p S_{\phi}^2 C_{\theta}) \\
           &\quad + (I_{y} - I_{z}) \phi_p C_{\phi}^2 C_{\theta} + I_{x} \psi_p S_{\theta} C_{\theta}, \\
    c_{33} &= (I_{y} - I_{z}) \phi_p C_{\phi} S_{\phi} C_{\theta}^2 - I_{y} \theta_p S_{\phi}^2 C_{\theta} S_{\theta} \\
           &\quad - I_{z} \theta_p C_{\phi}^2 C_{\theta} S_{\theta} + I_{x} \theta_p C_{\theta} S_{\theta}.
\end{aligned}
\]


\bibliography{biblio.bib}
\bibliographystyle{IEEEtran}

\vspace{11pt}

\begin{IEEEbiography}[{\includegraphics[width=1in,height=1.25in,clip,keepaspectratio]{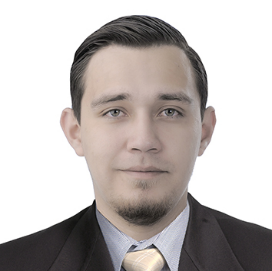}}]{Bryan S. Guevara} graduated in Mechatronic Engineering from the University of the Armed Forces - ESPE in 2018. In 2024, he successfully earned his master's degree in Control Systems Engineering from the National University of San Juan. Currently, he is a doctoral student in Control Systems Engineering at the same university, supported by the DAAD scholarship (German Academic Exchange Service) through the Funding Program: Third Country Programme Latin America, 2022. His areas of interest are: aerial robotics, dynamic systems modeling, and optimal control.
\end{IEEEbiography}

\vspace{11pt}

\begin{IEEEbiography}[{\includegraphics[width=1in,height=1.25in,clip,keepaspectratio]{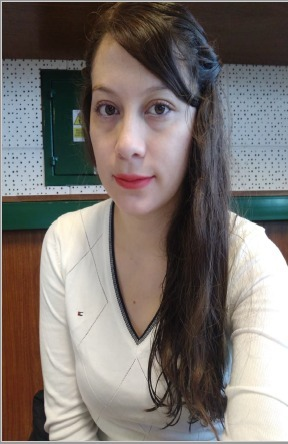}}]{Viviana Moya} is a Lecturer 
 and Researcher at the Universidad Internacional del Ecuador. In 2021, she has a Ph.D. in Control Systems Engineering from The National University of San Juan (UNSJ) in Argentina. Her doctoral studies were supported by a DAAD scholarship from Germany. Prior to this, Viviana completed her undergraduate education in Electronics and Control Engineering at the Escuela Politécnica Nacional (EPN) in Quito, Ecuador, in 2016. Her professional interests include teleoperation systems and automatic control. In recent years she has been focusing also on Artificial Intelligence and Computer Vision.
\end{IEEEbiography}

\vspace{11pt}

\begin{IEEEbiography}[{\includegraphics[width=1in,height=1.25in,clip,keepaspectratio]{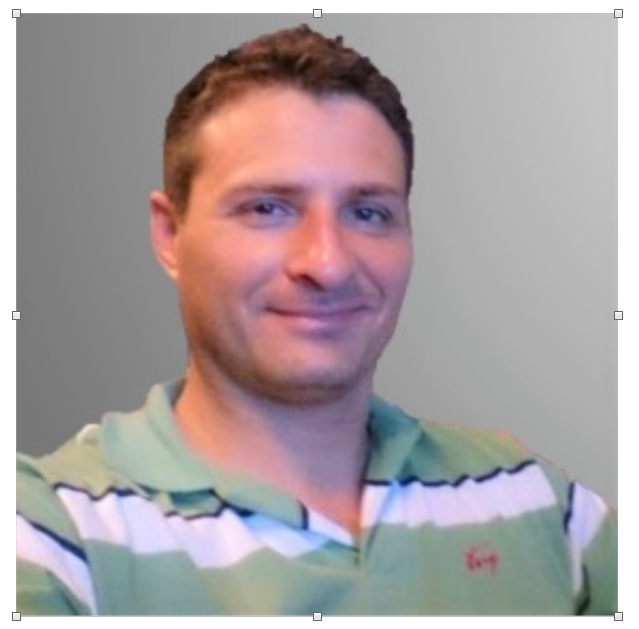}}]{Daniel C. Gandolfo} graduated in Electronic Engineering from the National University of San Juan (UNSJ), Argentina, in 2006. He has been working as automation engineer in the industry until 2009 and received the Ph.D. in Control Systems Engineering from the UNSJ in 2014. Currently, he is researcher of the Argentinean National Council for Scientific Research (CONICET), and an Associate Professor in the Institute of Automatics, UNSJ-CONICET, Argentina. His research interest included algorithms for management energy systems and optimal control strategies with application in unmanned aerial vehicles (UAV).
\end{IEEEbiography}

\vspace{11pt}

\begin{IEEEbiography}[{\includegraphics[width=1in,height=1.25in,clip,keepaspectratio]{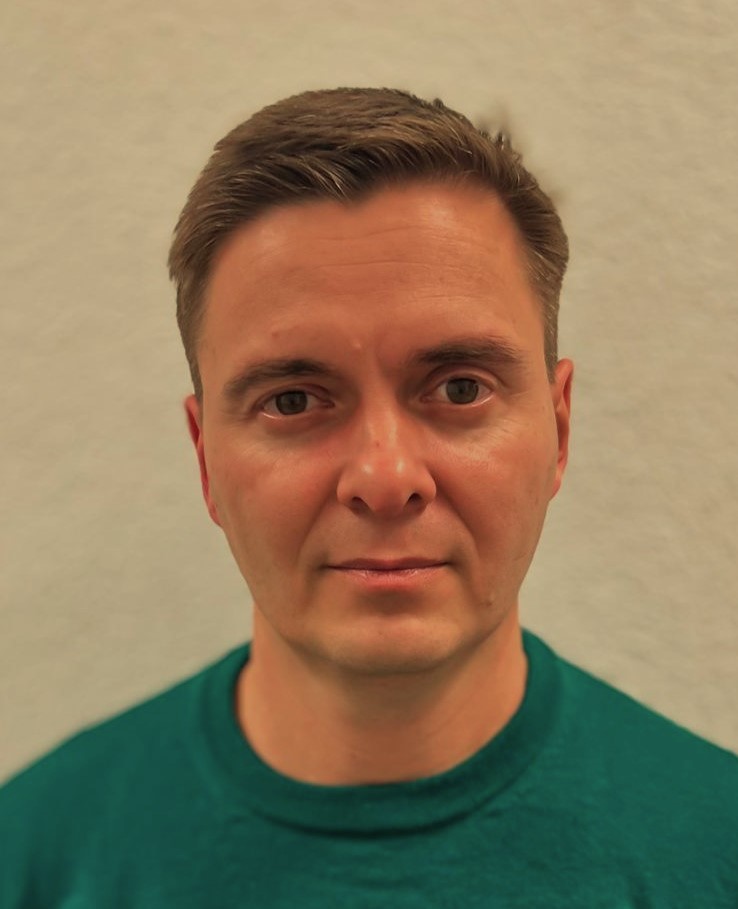}}]{Juan M. Toibero} graduated in Electronic Engineering from the National Technological University of Parana (Argentina) in 2002, received the Ph.D degree in Control Systems Engineering from the National University of San Juan (Argentina) in 2007. He is with the Instituto de Automática, National University of San Juan from 2002 as full professor and with CONICET from 2011 as full-time researcher.  His research interests are related to the automatic control of mobile robotic platforms. He's working currently with nonlinear control design, the inclusion of dynamic models and the applications of USVs (unmanned surface vessels) in environmental monitoring, surveillance and control. 
\end{IEEEbiography}

\vfill

\end{document}